\documentclass[useAMS,usenatbib]{mn2e}
\usepackage{psfig,subfigure}

\title[THE STRUCTURE OF BROWN DWARF CIRCUMSTELLAR DISKS]
 {THE STRUCTURE OF BROWN DWARF CIRCUMSTELLAR DISKS}

\author[Walker et al.]
 {Christina Walker$^1$, Kenneth Wood$^1$, C.J.~Lada$^2$,Thomas Robitaille$^1$,
  \newauthor 
J.E.~Bjorkman$^3$,Barbara Whitney$^4$\\ 
 $^1$School of Physics \& Astronomy, University of St. Andrews,
North Haugh, St Andrews, KY16 9SS, Scotland;\\
cw26@st-andrews.ac.uk, kw25@st-andrews.ac.uk, tr9@st-andrews.ac.uk\\
 $^2$Harvard-Smithsonian Center for Astrophysics, 
60 Garden Street, Cambridge, MA~02138; clada@cfa.harvard.edu\\
 $^3$Ritter Observatory, Department of Physics \& Astronomy, 
University of Toledo, Toledo, OH 43606; jon@astro.utoledo.edu\\
 $^4$Space Science Institute, 3100 Marine Street, Suite A353, 
Boulder, CO~80303; bwhitney@colorado.edu}

\date{Released 2003 Xxxxx XX}
\pagerange{\pageref{firstpage}--\pageref{lastpage}} \pubyear{2003}

\def\LaTeX{L\kern-.36em\raise.3ex\hbox{a}\kern-.15em
 T\kern-.1667em\lower.7ex\hbox{E}\kern-.125emX}

\begin{document}

\label{firstpage}

\maketitle

\begin{abstract}

We present synthetic spectra for circumstellar disks that are 
heated by radiation from a central brown dwarf.  Under the assumption 
of vertical hydrostatic equilibrium, our models yield scaleheights for 
brown dwarf disks in excess of three times those derived for classical 
T~Tauri (CTTS) disks.  If the near-IR excess emission observed from 
brown dwarfs is indeed due to circumstellar disks, then the large 
scaleheights we find could have a significant impact on the optical 
and near-IR detectability of such systems.  Our radiation transfer 
calculations show that such highly flared disks around brown dwarfs 
will result in a large fraction of obscured sources due to extinction 
of direct starlight by the disk over a wide range of sightlines.  
The obscured fraction for a 'typical' CTTS is less than 20\%. We 
show that the obscured fraction for brown dwarfs may be double that 
for CTTS, but this depends on stellar and disk mass.  We also comment on 
possible confusion in identifying brown dwarfs via color-magnitude 
diagrams: edge-on CTTS display similar colors and magnitudes as a 
face-on brown dwarf plus disk systems.

\end{abstract}


\section{Introduction}

Accumulating observational evidence indicates the presence of 
circumstellar disks around brown dwarfs, including near-IR 
(Oasa et al. 1999; Muench et al. 2001; Liu et al. 2003) 
and mid-IR excess emission (Comer\'{o}n et al. 1998; 2000), and 
H$\alpha$ signatures of accretion (Muzerolle et al. 2000).  The 
existence of circumstellar disks around brown dwarfs is an 
important discovery, since it may suggest that brown dwarfs 
form in a similar fashion to more massive T~Tauri stars (Shu, Adams,
\& Lizano 1987).  At the same time, data indicating 
significant masses and extents of circumstellar material may 
cause problems for brown dwarf formation scenarios where the 
low mass object is formed and subsequently ejected from multiple 
systems (Reipurth \& Clarke 2001).  

The observed spectral energy distributions (SEDs) and IR excess 
emission of some brown dwarfs have been modelled using flat and 
flared reprocessing disks (e.g., Natta \& Testi 2001; Testi et al. 
2002; Liu et al. 2003).  The SED models suggest brown dwarf disks 
are similar to those around CTTS. Models for the SEDs and scattered 
light images of CTTS require flared disks (e.g., Kenyon \& Hartmann 
1987; D'Alessio et al. 1999; Whitney \& Hartmann 1992; Burrows et al. 1996).  
For highly inclined flared disks direct starlight is blocked by the 
optically thick disk resulting in a fraction of sources that will 
appear very faint in the optical and near-IR.  Such faint or ``optically 
obscured'' sources may escape detection in magnitude limited surveys.  
The degree of disk flaring depends on the disk temperature structure 
and the mass of the central star, with the disk scaleheight 
$h\propto (T_d/M_\star)^{1/2}$ (Shakara \& Sunyaev 1973).  Therefore, 
low mass brown dwarfs may have more vertically extended disks than 
those around CTTS and, depending on the disk mass, the obscured 
fractions may be larger.  We will show that this may lead to confusion 
in discriminating between brown dwarfs and edge-on CTTS. 

In this paper we adopt the same working hypothesis as Natta \& Testi (2001) 
that brown dwarf disks are in vertical hydrostatic equilibrium with dust 
and gas well mixed throughout the disk.  Such disk models have been very 
successful in explaining the observed scattered light images and SEDs of CTTS.  
We extend our Monte Carlo radiative equilibrium code to calculate the 
structure of passively heated brown dwarf disks in vertical hydrostatic 
equilibrium.  Our Monte Carlo radiation transfer technique naturally 
includes scattered light and the inclination dependence of the SED, 
which allows us to investigate the effects of highly flared disks.  
We construct synthetic spectral energy distributions and colors for 
disks of different sizes and masses surrounding brown dwarfs of different 
masses and luminosities.  Our model SEDs enable us to determine to what 
extent observations in various spectral regions can diagnose disk parameters.  
Deriving disk parameters for large numbers of sources may help to 
discriminate brown dwarf formation mechanisms and whether they are 
different for dense and sparse star forming regions. 

The layout of the paper is as follows: \S2 outlines the ingredients of 
our models and the radiation transfer/disk density calculation, \S3 
presents disk structure models derived with our iterative technique, 
\S4 presents our model SEDs and color-color diagrams, \S5 compares our 
models with currently available observations of brown dwarf disks, and 
we summarize our results in \S6.

\section{Model Ingredients}

This study implements a number of extensions to the original Monte Carlo 
radiative equilibrium technique of Bjorkman \& Wood (2001).  These include 
a crude estimate of the inner disk radius (assumed to be at the dust 
destruction radius) and an improved temperature structure calculation.  
Our disks are not vertically isothermal or two-layered (Natta \& Testi, 2001); 
2D disk temperature structure is calculated 
in the MC simulation based on the technique described by Lucy (1999).  
Our code self-consistently determines the density structure of a passively 
heated disk in vertical hydrostatic equilibrium.  The extensions to the 
radiation transfer technique are described in greater detail in the Appendix. 

\subsection{Disk Structure Calculation}
Model SEDs are computed for a flared disk that is heated by radiation 
from a central brown dwarf.  We only consider passive disks, since disk 
heating from viscous accretion is negligible compared to stellar heating 
in low accretion rate systems (Muzerolle et al. 2000; D'Alessio et al. 
1999).  Our disks extend from the dust destruction radius to an outer 
radius of 100AU.  The disk is truncated sharply 
at its inner edge and there is no material between the inner edge and 
the star, equivalent to assuming material in this region is optically 
thin.  In our previous modelling of CTTS SEDs (Wood et al. 2002a, b; 
Schneider et al. 2003; Grosso et al. 2003) we adopted the following 
flared disk density structure (e.g., Shakara \& Sunyaev 1973)
\begin{equation}
\rho=\rho_0 \left ({R_\star\over{\varpi}}\right )^{\alpha}
\exp{ -{1\over 2} [z/h( \varpi )]^2  }  
\; ,
\end{equation}
where $\omega$ is the radial coordinate in the disk midplane and the 
scaleheight increases with radius,
$h=h_0\left ( {\varpi /{R_\star}} \right )^\beta$.  With the disk 
structure fixed we then calculate the temperature structure and 
emergent SED using the Monte Carlo radiative equilibrium technique 
of Bjorkman \& Wood (2001).

In this paper we adopt an iterative scheme to determine the disk density 
structure.  Having calculated the disk temperature structure via our Monte 
Carlo radiative equilibrium technique, we impose vertical hydrostatic 
equilibrium and solve 
\begin{equation}
{{dP}\over{dz}} = -\rho g_z \; .
\end{equation}
Here, $P=\rho c_s^2$ is the gas pressure, $c_s$ is the isothermal sound 
speed, and $g_z=\frac{GM_\star z}{\varpi^3}$ is the vertical component 
of gravity in the disk.  We make the usual thin disk assumptions and assume 
the disk is non self-gravitating (Pringle 1981).We impose the boundary 
condition that the disk surface density $\Sigma\sim \varpi^{-1}$, in 
accordance with the detailed disk structure models of D'Alessio et al. 
(1999).  Our simulations begin with the disk structure given by equation~1 
with $\alpha=2.25$, $\beta=1.25$, and we iterate to derive a self-consistent 
vertical density structure.  The density converges within three iterations.

In hydrostatic disk models, the disk scaleheight scales with radius as 
$h/ \varpi=c_s/v_c$, where $c_s^2=kT/\mu m_H$ and $v_c^2=GM_\star/ \varpi$ 
are the isothermal sound speed and circular velocity at $\varpi$ (e.g., 
Shakara \& Sunyaev 1973; Lynden-Bell \& Pringle 1974).  For CTTS, 
$h(100 {\rm AU})$ is in the range 7AU to 20AU as found from radiative 
and hydrostatic equilibrium models (D'Alessio et al. 1999) and from 
fitting SEDs and scattered light images of disks using equation~1 with 
$h_0$ as a free parameter (Burrows et al. 1996; Stapelfeldt et al. 1998; 
Grosso et al. 2003; Schneider et al. 2003).  However, for disks around 
brown dwarfs the scaleheights may be larger due to the smaller circular 
velocity of these low mass objects.  If brown dwarf disks are indeed more 
vertically extended, then there may be a larger fraction of obscured 
brown dwarfs compared with CTTS.  Our SED calculations enable us to address 
this issue.

\subsection{Dust Parameters and Model Atmospheres}
The circumstellar dust opacity and scattering properties are taken to be 
those of the dust size distribution we adopted for modelling the SEDs of 
HH~30~IRS and GM~Aur (Wood et al. 2002a; Schneider et al. 2003; Rice et al. 
2003).  This dust model has a larger average grain size and a shallower 
wavelength dependent opacity than ISM dust models (e.g., Mathis, Rumpl, 
\& Nordsieck 1977; Kim, Martin, \& Hendry 1994).  There is much observational 
evidence for large grains and a shallow wavelength dependent opacity in 
T~Tauri disks (e.g., Beckwith et al. 1990; Beckwith \& Sargent 1991; 
D'Alessio et al. 2000; Cotera et al. 2001; Wood et al. 2002a).  The larger 
grain dust model we adopt does not exhibit strong silicate features 
(see Wood et al. 2002a).

The input stellar spectra for the brown dwarf models are the BD\_Dusty
model atmospheres presented by Allard et al. (2001),  with $\log g = 3.5$ 
and effective temperatures of $T_\star = 2200$~K, 2600~K and 2800~K.  For 
CTTS models we use a 4000~K Kurucz model atmosphere (Kurucz 1994).

\subsection{Parameter Space}
We construct radiative and hydrostatic equilibrium models for brown dwarf 
systems with the range of stellar and circumstellar disk parameters given 
in Table 1.  The stellar mass range of 
$0.01M_\odot\le M_\star \le 0.08M_\odot$ covers objects from the hydrogen 
burning limit down to the lower limit for brown dwarfs as identified via 
color-magnitude diagrams by Muench et al. (2001).  The corresponding stellar 
radii and temperatures yield models representative of 1Myr old systems from 
the evolutionary tracks of Baraffe et al. (2002).  For each set of stellar 
parameters, disk to star mass ratios of log($M_d/M_\star$)=-1, -2 and -3 
are initially considered.  The disk mass $M_d$ refers to the total disk mass 
of dust and gas.  As with our previous work, this mass does not include very 
large particles such as rocks or planetesimals and is therefore a lower 
limit.  We compare our resulting brown dwarf disk structures with those of 
disks around a typical CTTS with $M_\star = 0.5M_\odot$, $R_\star = 
2R_\odot$, and $T_\star = 4000$~K (e.g., Kenyon \& Hartmann 1995; D'Alessio 
et al. 1999).

\section{Disk Structure Models}

At the end of our iterative procedure (described in the Appendix) the 
outputs of our code are the disk density and temperature structure 
and the emergent SED.  All our brown dwarf models have Toomre parameter, Q, 
$>$ 1, throughout their disks,  so the thin disk assumption implicit in 
our models is still valid (e.g. D'Alessio et al. 1999).  Fig.~1 shows 
scaleheights and mid-plane temperatures for disks of various masses 
illuminated by stars of different mass.  The full disk structure is now 
calculated, however we choose to define scaleheight using the mid-plane 
temperature (see Appendix).  For comparison we show the scaleheight and 
mid-plane temperature for a CTTS illuminating disks of the same mass ratio 
as in the brown dwarf models.  The scaleheights of the CTTS disks are 
$h(100\,{\rm AU})\sim 15$~AU, in agreement with the simulations of 
D'Alessio et al. (1999).  The brown dwarf disks have scaleheights 
significantly in excess of those obtained for CTTS, with $h(100\,{\rm AU})$ 
ranging from just over 20~AU for $M_\star=0.08M_\odot$ to almost 
60~AU for $M_\star=0.01M_\odot$. 

Our temperature calculations in Fig.~1 for brown dwarf disks show that 
$T (100\,{\rm AU}) \sim 10$~K with little variation among the models.  
The stellar mass therefore predominantly controls the disk scaleheights.  
The brown dwarf models show disk scaleheights up to three 
times larger than for comparative disks illuminated by a CTTS.  Such 
large scaleheights will result in a large range of viewing angles for 
which direct starlight will be extincted by the disk.  The effects of 
large scaleheights on the SED and colors are discussed in \$4.

The extended nature of the brown dwarf disks is also clear in Fig~2 which 
shows K-band scattered light images of disks viewed at an inclination of 
$85^o$ from face-on.  As with CTTS models (Wood et al. 1998), the dust lane 
narrows with decreasing disk mass and the central source becomes 
increasingly more visible.  Hence it seems that the detection 
of low mass disks via scattered light may only be possible for edge-on 
systems or if coronographic techniques are used to block the starlight.

\section{Model Spectra and Colors}

In addition to calculating the disk structure, our radiation transfer code 
outputs the SED for a range of viewing angles.  This section shows SEDs 
and color-color diagrams that illustrate the main features of our models.  
With the Monte Carlo technique it is straightforward to determine the 
contributions to the SED of stellar, scattered, and thermally reprocessed 
photons (Wood et al. 2002a).  We utilize this capability to determine the 
relative importance of the scattered light contribution to the SEDs and 
colors.

\subsection{SEDs of Face-On Disks}

Our brown dwarf model SEDs have similar spectral characteristics to 
those of CTTS disks (e.g., Wood et al. 2002b).  Fig.~3 shows SEDs of face-on 
disks for a range of star and disk parameters.  Face-on covers 
0 $\rightarrow$ $18^o$ due to binning of the photons in the Monte Carlo 
code.  The dependence of SED on disk mass is readily evident and, as with 
CTTS, observations at long wavelengths provide the best diagnostics of disk 
mass.

As commented by Natta \& Testi (2001) it is difficult to produce significant 
near-IR excesses for brown dwarfs because the stellar spectrum peaks at 
longer wavelengths than CTTS and can therefore dominate the disk thermal 
emission.  At longer wavelengths however Fig.~3 shows that our brown 
dwarf models are capable of producing varying degrees of IR excess emission. 
As stellar mass decreases, scaleheights increase allowing the disk to 
intercept, scatter, and thermally reprocess more stellar radiation, 
which in turn gives rise to increasingly large IR excesses.
 
Fig.~4 shows the relative contribution of stellar, scattered, and thermal 
disk radiation for the highly flared $M_\star=0.01M_\odot$, 
$T_\star = 2200$~K brown dwarf disk system with $\log(M_d/M_\star)=-1$ and 
includes a CTTS model for comparison.  Scattered light makes little 
contribution to face-on models, but it can account for up to 90\% of 
K-band flux as disks become more inclined (see Wood et al. 2002b, Fig. 9).  
The importance of including scattered light will be highlighted in \S4.3. 

Recent work on brown dwarf formation suggests that many brown dwarfs are 
ejected from multiple systems and that any circumstellar disks that 
survive the ejection will be very small.  In the numerical simulations of 
Bate et al. (2003), no disks survive around ejected brown dwarfs 
down to their simulation resolution of $\sim 10$~AU.  We have computed SEDs 
for disks of constant mass, but varying $R_d$ in the range 10AU - 200AU. 
Because $M_d$ was held constant in these models, smaller $R_d$ yields larger 
optical depths.  The SEDs are mostly unaltered as $R_d$ changes apart from 
some variation at far-IR/sub-mm wavelengths.  We conclude that it would 
be very difficult to determine disk radii from SED data alone and more 
stringent tests of the small disks prediction of Bate et al. (2003) 
will require high resolution imaging to resolve the disks via their 
scattered light and thermal emission (see also, 
Beckwith et al. 1990; Chiang et al. 2001).  

\subsection{Near-IR Color-Color Diagrams}

By far the most popular technique of identifying circumstellar disks 
is to identify sources with near-IR excess emission in color-color 
diagrams (e.g., Lada \& Adams 1992; Rebull et al 2002).  It was through 
near-IR color-magnitude and color-color diagrams that Meunch et al. 
(2001) and Liu et al. (2003) identified many candidate brown dwarfs that 
exhibit the tell-tale IR excess emission indicative of circumstellar disks.  

All colors we present are relative to Vega and are computed using 2MASS JHK 
and UKIRT L filter transparency curves.  The BD\_Dusty model atmospheres 
that we use have near-IR colors that are bluer than observations of the 
corresponding spectral type (e.g., Bessell 
\& Brett 1988; Kirkpatrick et al. 2000).  What is important is the 
relative color of our models (e.g., $[H-K] - [H-K]_\star$) and the 
underlying stellar spectrum does not affect this.  As we ultimately 
compare our models with observations, we have used a similar approach 
to Liu et al. (2003) and applied a color offset to the models so that 
the model stellar colors match observations.  We adopt spectral types of 
M9.5, M8.5 and M6 for our 2200, 2400 and 2800~K models respectively.  There 
is no well defined temperature scale for M dwarfs and so classifications 
were chosen on consideration of observations and discussion by Luhman 
(1999), Pavlenko et al (2000) and Dahn et al (2002).  We shift the 
stellar colors of our models to match the field M dwarf locus taken from 
Bessell \& Brett (1988) and average colors from Kirkpatrick et al. (2000).  
This results in the following offsets for the M9.5, M8.5 and M6 fits:
$\Delta(J-H) = 0.34,0.23,0.10$, $\Delta(H-K) = 0.12,0.06,0.00$. 
No shift in K-L is applied. 

Fig.~5 shows $JHK$ and $JHKL$ color-color diagrams for our model disks 
viewed face-on, and following the afore mentioned adjustments.  In general, 
excess emission is more readily detected at long wavelengths 
(e.g. Haisch, Lada, \& Lada 2000, Natta \& Testi 2001) and this is again 
seen here with models showing larger excesses at $K-L$ than at $J-H$ or 
$H-K$.  The trend of our models is that the more massive and more flared disks 
exhibit the largest IR excesses.  Inclination effects yield a spread in 
color-color diagrams and we explore this in the next section.

\subsection{Inclination, Scattered Light, and Obscured Fractions}

For highly inclined CTTS, direct starlight is blocked by the optically 
thick disk and such systems will be very faint in the optical and 
near-IR (e.g., D'Alessio et al. 1999; Wood et al. 2002a, b).  Compared 
to CTTS, the larger disk scaleheights we derive for the brown dwarf models 
will result in a larger fraction of viewing angles over which the 
central starlight is blocked by the disk.  For the purpose of this study 
we define an ``obscured source'' to be one where the near-IR flux 
is at least three magnitudes fainter than the corresponding face-on source.  
The obscured fraction therefore depends on the disk size, mass, and 
scaleheight.  For CTTS, the obscured fraction is around 20\% (D'Alessio et 
al. 1999; Wood et al. 2002b) for disks of $M_d\sim 10^{-3}M_\odot$.  

Figure~6 shows the inclination effect on the SEDs for various brown 
dwarf disk models.  The SEDs are shown for ten viewing angles evenly 
spaced in $\cos i$, so that each curve represents 10\% of sources by 
number if we assume sources are randomly distributed in inclination angle.  
We find obscured fractions from $20\%$ to $60\%$ with highly inclined 
sources only detected in the near-IR via scattered light and weak 
thermal emission.  The largest obscured fraction occurs for the lowest 
stellar mass of $M_\star=0.01M_\odot$ with log($M_d/M_\star$)=-1 and 
$T_\star = 2200$~K.  The smallest obscured fraction occurs for the highest 
stellar mass of $M_\star=0.08M_\odot$ with log($M_d/M_\star$)=-3 and 
$T_\star=2800$~K.  

Studies of the initial mass function in Trapezium, $\rho$ Ophiucus and 
IC348 show a relatively flat distribution over the range 
$0.08 \le M_\star(M_\odot) \le 0.04$ and then a sharp fall off below this
(Luhman 2000; Muench et al. 2002).  If the IMF is flat and the fall-off 
due to small number statistics then within a young cluster population up to
55\% of brown dwarf candidates, as defined by our parameter range, may be 
obscured.  This is an upper limit produced using maximum obscuration 
fractions for each stellar mass assuming a disk to stellar mass ratio of 
log($M_d/M_\star$)=-1.  For a declining IMF, and a distribution of disk 
masses, the obscured fraction will be less.  Within our parameter range 
a minimum of $20\%$ of sources are likely to be obscured regardless of 
stellar mass distribution and assuming disk to stellar mass ratios of 
log($M_d/M_\star$)=-3.   

The relatively low luminosity of brown dwarfs and the increased obscuration 
due to highly flared disks may present detection problems.  At a 
distance of 150pc (as used in Fig.~6) it would be possible to detect 
some obscured sources in the K band assuming a sensitivity limit of 
16.5 mags.  In the absence of high resolution imaging however these sources 
may be incorrectly identified as low luminosity systems.  A three 
fold increase in distance would be sufficient to make all obscured 
sources undetectable at this sensitivity limit.

Figure~7 shows the inclination dependence in the brown dwarf $JHKL$ 
color-color diagrams.  Relative colors are plotted for ten inclinations with 
the change in color at each inclination indicated by an arrow.  Similar 
to the behaviour observed by Kenyon et al. (1993) and Whitney et al. 
(1997) we see a loop in the color-color plane with inclination.  
Starting from face-on, the sources generally get redder with increasing 
inclination and then loop around and end up with edge-on sources being 
slightly bluer than face-on, but still redder than the intrinsic stellar 
colors.  Edge-on sources are seen almost entirely via scattered light. 
Note that these are slightly redder than the star because the scattered 
light, which is relatively blue, suffers extinction and becomes somewhat 
reddened.  This trend is seen in all of the models.

Figure~8 contains data for the same model as in Fig.~7, but also shows the 
change in color with inclination if scattered light is ignored.  The removal 
of scattered light makes the colors much redder, with the effect being 
particularly significant at moderate to high inclinations.  This emphasizes 
the importance of including scattering when creating and studying models 
of such systems.  

\subsection{CTTS/Brown Dwarf Confusion}
When only unresolved photometry is available our models show that 
edge-on CTTS could be mistaken as brown dwarfs.  CTTS have edge-on 
flux levels that are comparable to face-on brown dwarfs and similar 
colors.  Muench et al. (2001) identified sources within the Trapezium 
cluster with $13.5\la H\la 17.5$ as candidate brown dwarfs.  They note 
that 21 of their 109 brown dwarf candidates are coincident with optically 
resolved proplyds (Bally, O'Dell, \& McCaughrean 2000; O'Dell \& Wong
1996) and 21\% of the candidates that exhibit IR excess, indicative of 
circumstellar disks, are represented by these proplyds.  In the absence 
of high resolution imaging the task of identifying faint sources such 
as brown dwarfs may be problematic.  If no central star is seen then 
these sources could be edge-on CTTS that happen to have the same magnitude 
and colors as a pole-on brown dwarf and disk.  This confusion could lead 
to an overestimation of brown dwarf numbers.   

\section{Comparison to Observations}

This section compares our synthetic models and published observations of 
suspected brown dwarf disks.  For the Chameleon cluster the SED data is 
taken from Comer\'{o}n et al (2000) and Apai et al (2002);  $\rho$ 
Ophiucus data comes from Barsony et al. (1997), Comer\'{o}n et al (1998), 
Bontemps et al (2001) and Natta et al (2002).  JHKL data is taken from the 
above papers along with Kirkpatrick et al (2000) and Liu et al.(2003).  
All near-IR photometry has been converted to the 2MASS system 
(Carpenter, 2001).

\subsection{Spectral Energy Distributions}

Figure~9 shows flared disk model fits to the SED data for candidate 
brown dwarfs in the $\rho$ Ophiucus and Chameleon star clusters.  Table 2. 
contains details of the model parameters used to produce the fits.  We used 
stellar parameters from  Natta \& Testi (2001) and Natta et al (2002) 
as starting points for each of our models. 

Natta \& Testi (2001) modelled Cha H$\alpha$1, 2, \& 9 using a flared disk 
model.  Their models produced a successful fit in the MIR region of the 
spectrum and predicted a strong 10$\mu$m silicate emission feature.  
Apai et al (2002) later made observations of Cha H$\alpha$2 at 9.8 and 
11.9 $\mu$m and did not detect the silicate feature.  They presented  
an optically thick flat disk model which produced no silicate feature.  
The SEDs of $\rho$ Oph sources have been modelled by Natta et al. (2002) 
and they found indications that as many as eight of these stars may have 
flat disks.  

Our models open up the possibility that the absence of a silicate feature 
may be explained with larger circumstellar dust grains.  In addition, low 
mass disks may fit SEDs previously modelled with flat disks.  As Fig.~9 
demonstrates, it is possible to fit the observed data for all sources with 
a flared disk geometry.  The use of larger grains naturally suppresses the 
silicate feature which has been shown to be missing from the Chameleon data 
and low mass disks of $10^{-5}M_\odot$ and $10^{-7}M_\odot$ allow us to fit 
the IR data of the candidates where flat disks were previously suspected.  
We note that many of these fits have disk to stellar mass ratios outside the 
typical range of $-1 \le log(M_d/M_\star) \le -3$ (Natta et al. 2000; 
Klein et al. 2003) and flat disks (Natta et al. 2002) remain a possibility.  

Another alternative, testable with long wavelength observations, is that 
steeper surface density profiles can also be used to fit the data with 
higher mass disks.  In Fig.~10, ISO\#030 has been modelled using both 
surface density $\Sigma\sim \varpi^{-1}$ and $\Sigma\sim \varpi^{-2}$.  
Using $\Sigma\sim \varpi^{-2}$ allows us to fit the data with a disk eight 
times more massive disk than used in the $\Sigma\sim \varpi^{-1}$ case.  
Both models fit the data well in the NIR/MIR, but are quite different 
in the FIR. Long wavelength observations would help to discriminate 
between flat disk, low mass flared disks and steeper surface density disk 
models.  
  
If lower mass flared models are 
representative of disks in brown dwarf populations, as opposed to higher 
mass disks, then problems with obscuration may not be as significant as 
suggested in \S4.3.  Equally flat disks do not result in severe 
obscuration of the central star unless at very high inclinations.  

Figure~11 shows the derived scaleheights for the disks that we used to 
model the observed SEDs of Fig.~9 and Fig.~10.  
This illustrates the range of disk 
structures that can produce fits to the observed data.  In each plot the 
scaleheight of a model CTTS of corresponding disk to stellar mass ratio 
is presented as a comparison.  For these models we find scaleheights up 
to three times that of the corresponding CTTS.

\subsection{JHKL colors}
Figure~12 shows $JHKL$ plots of our face-on models and published data.  
Following the adjustments discussed in \S4.2, Fig.~11 shows that our 
models (if reddening were included) can reproduce the observed spread 
in colors of suspected brown dwarf disk systems.  Including all inclinations 
(Fig.~13) allows for the redder colors of inclined disks and produces a 
spread in the $JHKL$ plots that is in very good agreement with the observed 
colors.  

Figure~13 also shows the CTTS locus taken from Meyer et al. (1997).  
This again demonstrates that there is an overlap between CTTS and 
brown dwarf colors which may lead to incorrect identification of sources 
if only color-magnitude data is available.

\section{Summary}

We have presented model SEDs and color-color diagrams for brown dwarf 
disks.  The main assumptions in our models are that the disks are in 
vertical hydrostatic equilibrium with dust and gas well mixed throughout.  
Our models are self-consistent and employ an iterative procedure to 
determine the hydrostatic density structure for passively heated disks.  
Compared to CTTS, brown dwarf disks have larger scaleheights due to the 
lower mass of the central star.  In some cases the scaleheights of brown 
dwarf disks are more than three times larger than for the same disk to 
stellar mass ratio for a CTTS.  The larger scaleheights result in more 
inclinations over which the direct stellar radiation is blocked or obscured 
by the flared disk.  The fraction of optically obscured systems depends 
on the stellar mass and disk optical depth and in our models is in the 
range $20\%\le f_{\rm obs}\le 60\%$.  For a typical CTTS about 20\% of 
sources will be optically obscured.  

If, as our models suggest, brown dwarf disks are highly flared, detection 
of brown dwarf disk systems will be biased towards face-on systems.  We 
also show that without direct imaging or spectroscopic identification, it 
will be difficult to distinguish between edge-on CTTS and face-on brown 
dwarfs.  Color-color diagrams show that edge-on sources, which are only 
detected in the optical/near-IR via scattered light, have similar colors 
to face-on sources.  This may lead to incorrect identification of sources.  
In particular, we find that an edge-on CTTS will have similar near-IR 
magnitudes and colors as face-on brown dwarf disk systems. 

We compare our synthetic models to SED and color-color observations of 
suspected brown dwarfs and show that flared disks of varying mass can 
account for the observed SEDs and colors.  Our adopted circumstellar 
dust model naturally suppresses the 10$\mu$m silicate feature that is 
absent in the observations of Cha H$\alpha$2.  Long wavelength observations 
are required to discriminate between our flared disk models and alternative 
flat disk models that have been proposed for some sources.  

We acknowledge financial support from a UK PPARC Studentship (CW); 
UK PPARC Advanced Fellowship (KW); NASA's Long Term Space Astrophysics 
Research Program, NAG5~8412 (BW), NAG5~8794 (JEB); 
the National Science Foundation, AST~9909966 (BW), 
AST~9819928 (JEB).

\newpage

\renewcommand{\theequation}{A-\arabic{equation}}
\setcounter{equation}{0}  
\begin{center}
\section*{APPENDIX}  
\end{center}

\subsection*{MONTE CARLO RADIATION TRANSFER CODE DEVELOPMENTS}
\
\section*{A1. DENSITY STRUCTURE}
In this study we determine the structure of circumstellar disks
based on the assumption that the disk is in vertical
hydrostatic equilibrium with dust and gas well mixed.  We therefore
solve the hydrostatic equilibrium equation
\begin{equation}
{{dP}\over{dz}} = -\rho g_z\; ,
\end{equation}
where $P$ is the pressure, $\rho$ is the density, and $g_z$ is
the vertical component of gravity in the disk.  Conservation of mass
in the disk is enforced by keeping the radial dependence of the surface 
density.

The hydrostatic equation has an analytic solution if the disk temperature
is assumed to be vertically isothermal at any given cylindrical radius,
$\varpi$.  
Using the mid-plane temperature, $T(\varpi)$, the disk density has a
Gaussian distribution about the midplane with scaleheight
\begin{equation}
h(\varpi)=\left(\frac{k\, T(\varpi)\,\varpi^3}
{GM_\star \mu\, m_H}\right)^{\frac{1}{2}}\; ,
\end{equation}
where $k$ and $G$ are the Boltzmann and Newton
constants, $m_{H}$ is the mass of hydrogen and $\mu$ is the
molecular weight of disk material and is taken to be $\mu=2.3$
for a molecular hydrogen/helium combination.

In order to solve for the density numerically, we approximate the
integral of equation A-1 to a sum of finite contributions.
Using the equation of state, $P=\rho c_s^2$, where
$c_s^2={kT}/{\mu m_h}$ is the local sound speed squared, this leads to
\begin{equation}
\ln\left({{\rho}\over{\rho_0}}\right) =
{-\sum{\frac{1}{T}\left(\frac{dT}{dz}+
\frac{g_{z} \mu\, m_{H}}{k}\right){\Delta{z}}}}
\end{equation}
which can be solved using
\begin{equation}
{{dT}\over{dz}}=\cos\theta\frac{dT}{dr}-\frac{\sin\theta}{r}\frac{dT}{d\theta}
\end{equation}
In the discretization of the disk density we use a spherical polar grid
(Whitney \& Wolf 2002) throughout which we calculate $g_z$ at the midpoint
of each cell.  The cell temperature determined from our radiative
equilibrium calculation is assumed to be uniform within each cell and
$\Delta$z is the incremental distance through each cell which lies 
directly below grid centre $(r,\theta,\phi)$. It is therefore
possible to obtain values of ${\rho}/{\rho_{0}}$ for each grid cell.

We assume the disk surface density has the form,
$\Sigma(r)=\Sigma_0(\varpi/{R})^{-1}$, which agrees with the disk 
structure models of D'Alessio et al. (1999).  Since total disk mass is given by
\begin{equation}
M_d=\int_{Rmin}^{Rmax}\Sigma(\varpi)2\pi\,\varpi\, {\rm d}\varpi\; ,
\end{equation}
we can solve for $\Sigma_0$ and in turn get $\Sigma(\varpi)$.  
We then normalize $\rho_{0}$ so that surface density is a
constant,
\begin{equation}
\rho_0(\varpi)=\frac{\Sigma_0\left({R}/{\varpi}\right)}
{\sum{\rho}/{\rho_0}\,{\Delta z}}
\end{equation}
For each cell ${\rho}/{\rho_{0}}$ and the cylindrical radius, $\varpi$,
are known, so we may therefore determine the density in each cell.

\section*{A2. TEMPERATURE STRUCTURE}

In order to determine the density structure we require an accurate
calculation of the disk temperature structure.  While the Bjorkman \& Wood
(2001) technique yields accurate SEDs, we found that the temperature
calculation was too noisy for use in our density calculation.  Increasing
the number of photon energy packets yields a smoother temperature
structure at the cost of a large increase in CPU time.  We have therefore
implemented the temperature calculation technique of Lucy (1999), which is
based on using an estimator for the mean intensity of the radiation field.
Integrating this technique into our Monte Carlo code leads to a higher
signal-to-noise in the disk temperature determination with fewer photon
packets.  What follows is an outline of the procedure we use and we refer
the reader to Lucy (1999) for more details.

Provided a system is in radiative equilibrium, the rate at which matter
absorbs energy from the radiation field is balanced by the rate 
at which matter emits energy, $\dot{A}=\dot{E}$ or,
\begin{equation}
4\pi \int_0^{\infty}\rho(1-a_\nu)\kappa_{\nu}J_{\nu}\,{\rm d}\nu=
4\pi \int_0^{\infty}\rho(1-a_\nu)\kappa_{\nu}B_{\nu}\,{\rm d}\nu\
\end{equation}
As discussed by Lucy (1999) the mean intensity, $J_\nu$ and therefore
heating is proportional to the photon path lengths, $l$, through the cells 
yielding,
\begin{equation}
{\dot{A}}=\frac{\epsilon}{\delta{t}\delta{V}}
\sum{l}\rho(1-a_\nu)\kappa_{\nu}\; ,
\end{equation}
where $\delta{t}$ is the cell simulation time, 
$\delta{V}$ is the cell volume, 
$a_\nu$ is the scattering albedo, and $\kappa_\nu$ is the total opacity in 
$cm^2/g$.  The energy of each photon packet is $\epsilon=L\,\delta t/N$, where 
$L$ is the source luminosity and $N$ is the number of Monte Carlo 
photon packets used in our simulation.  
The expression for the rate at which matter emits energy can also be  
simplified to:
\begin{equation}
\dot{E}=4\pi\rho\kappa_P\,B(T)
\end{equation}
where $\kappa_P$ is the Planck mean absorption coefficient and 
$B(T)=\sigma\,T^4/\pi$ is the integrated Planck function.  
Equating A-7 and A-8 leads to the following expression for temperature:
\begin{equation}
T^4=\left(\frac{\dot{A}}{4\kappa_{p}(T)\sigma}\right)\; .
\end{equation}
Since $\kappa_{P}(T)$ is a function of temperature we solve this 
equation iteratively using pre-tabulated values of $\kappa_{P}(T)$.

Our code uses the Bjorkman \& Wood (2001) 
technique for reprocessing photon packets and our modification of 
Lucy's (1999) pathlength technique to determine the cell temperature.  
Since we assume the opacity is not a function of temperature, we do 
not need to iterate to determine the temperature structure, as discussed 
in Bjorkman \& Wood (2001).

\section*{A3. DUST DESTRUCTION}

Model disks used in this study extend from a sharply cut-off inner radius 
out to a specified distance. The inner radius is defined by 
the dust destruction temperature which we take to be 1600~K.  
We make the simplifying assumption that the dust destruction 
radius is independent of latitude in the disk.  
The determination of the dust destruction radius is carried out after the 
temperature calculation and involves a nested loop that counts how many 
grid cells at each radius have temperatures below 1600K.  If this number is 
outside some specified range then the inner radius is shifted either towards 
or away from the central star.  The disk density is then re-gridded and the 
temperature calculation/dust destruction radius determination repeated.  
This continues until a stable radius is established. Once the inner 
radius is fixed the program starts to iteratively solve for density 
as described above.

\section*{A4. ITERATIVE PROCEDURE}

The program self-consistently solves for density using an iterative 
procedure.  For the first iteration the analytic density structure of 
Eq.~1 is assumed and this allows an initial temperature structure to be 
found.  On the next iteration or at the point at which a suitable 
dust destruction radius has been established and a new grid set-up, 
this temperature structure is used to determine a density structure 
using the numerical density technique. This density replaces 
the analytic density structure and the next iteration begins.  
The procedure continues until the temperature and density structure 
converges, typically within three iterations.

\newpage

\begin{figure}
\centerline{\psfig{figure=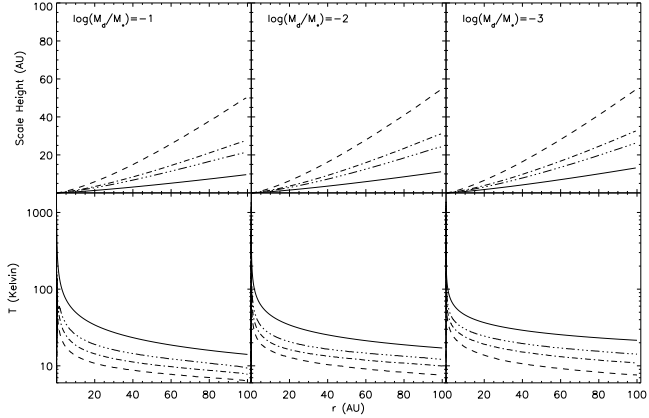,angle=0,width=3.5in}}
\caption[]{Upper: Scaleheights of brown dwarf models compared 
to CTTS models with matching disk mass. Lower: Mid-plane temperatures.  
In each plot the dashed, dot-dashed and triple dot dashed lines 
represent central stars of mass $0.01M_\odot$,$0.04M_\odot$ and $0.08M_\odot$ 
respectively and the solid line represents the CTTS model.  
The disk to stellar mass ratio is indicated in each panel.}
\end{figure}

\newpage


\begin{figure}
\centerline{\psfig{figure=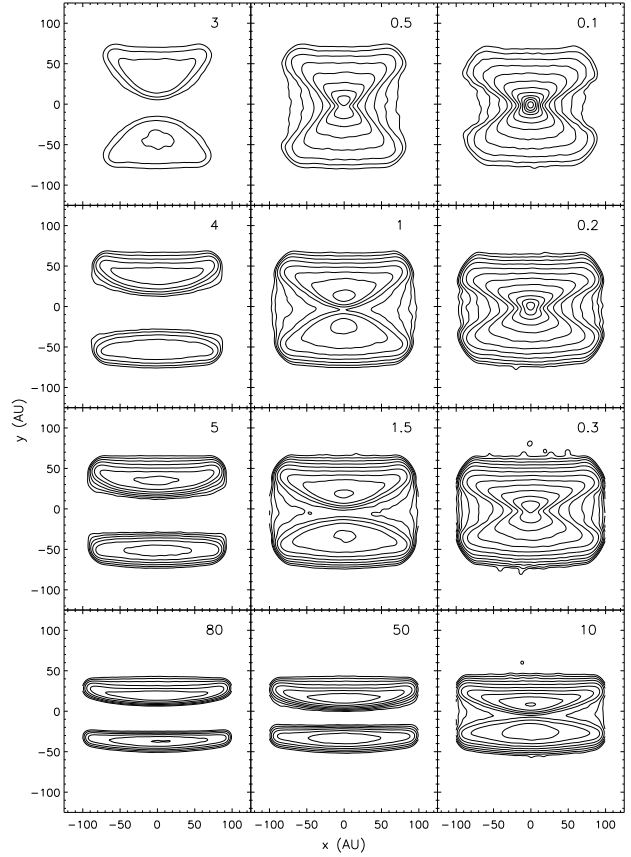,angle=0,width=3.5in}}
\caption[]{K-band contour plots of 9 brown dwarf models and 
3 comparison CTTS models. In declining order each row represents 
models of stellar mass 0.01, 0.04, and 0.08 $M_\odot$ and the bottom row 
is the CTTS model. Moving from left to right each column 
represents models with log($M_d/M_\star$)=-1, -2, and -3.  
The number in each panel is the lowest contour level in $\mu$Jy/$arcsec^2$, 
assuming source is at 150pc.  Contours increase in 1 mag intervals.}
\end{figure}

\newpage

\begin{figure}
\centerline{\psfig{figure=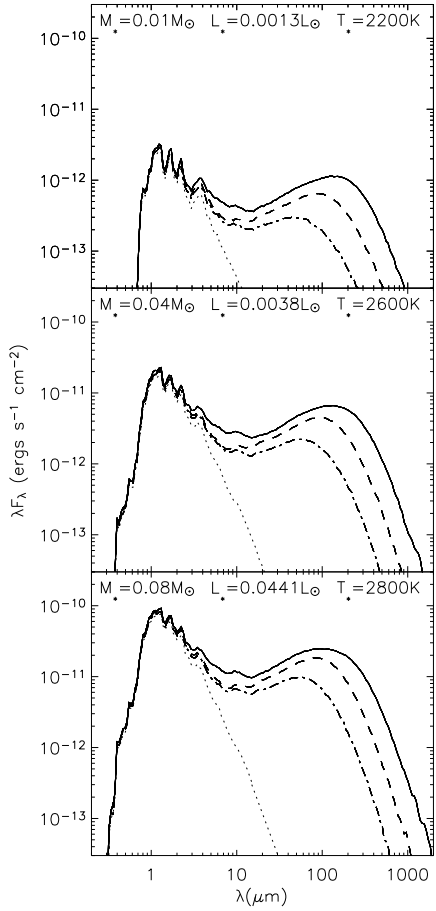,angle=0,width=2.5in}}
\caption{Face-on model SEDs showing the effects of varying stellar properties 
and disk mass. Each plot contains three separate SEDs for 
log($M_d/M_\star$)=-1 (solid line), -2 (dashed line) and -3 (dot-dash line).  
The dotted line represents the input stellar spectrum.  
Other parameters are as described in the text.}
\end{figure}

\newpage

\begin{figure}
\centerline{\psfig{figure=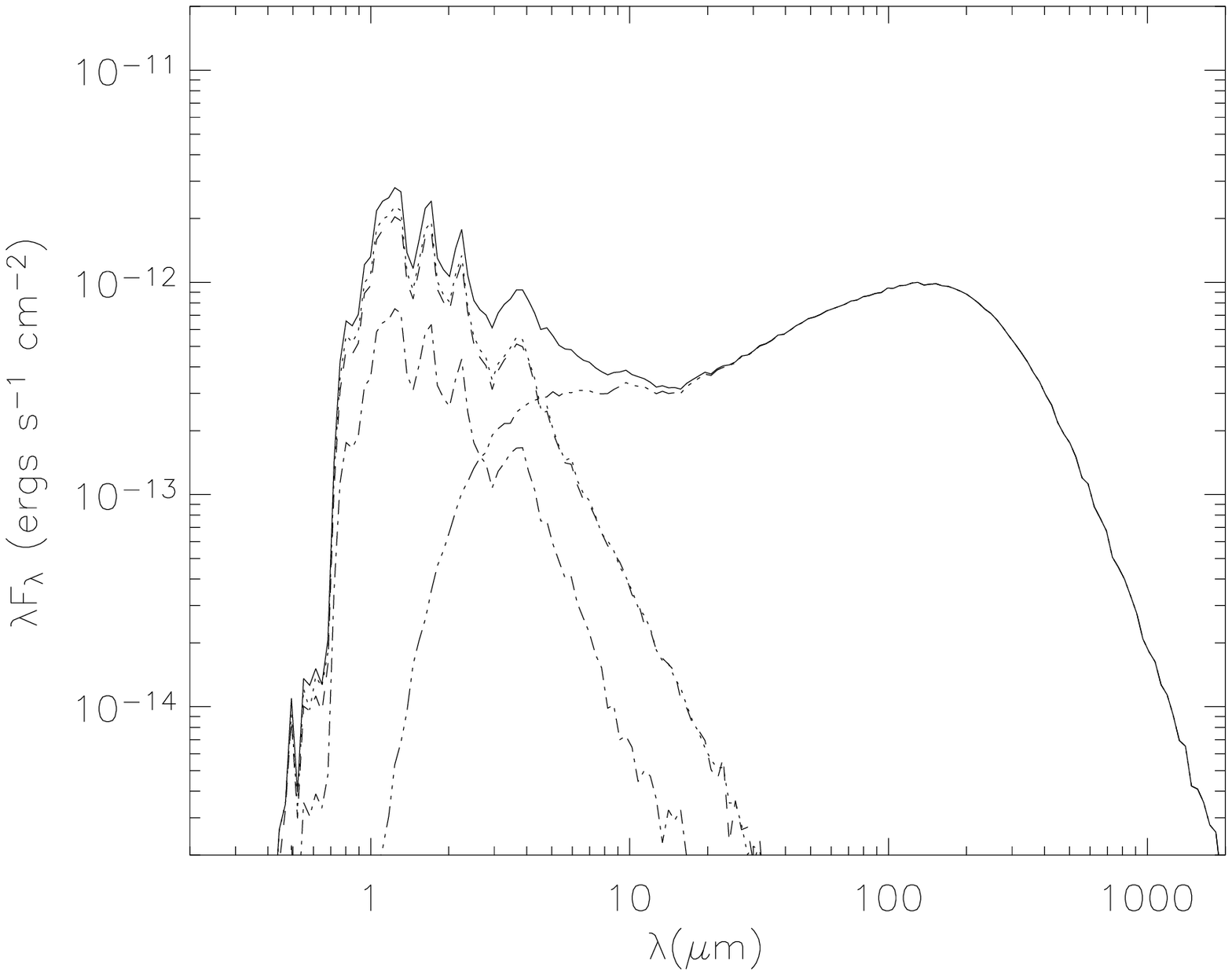,angle=0,width=2.5in}}
\centerline{\psfig{figure=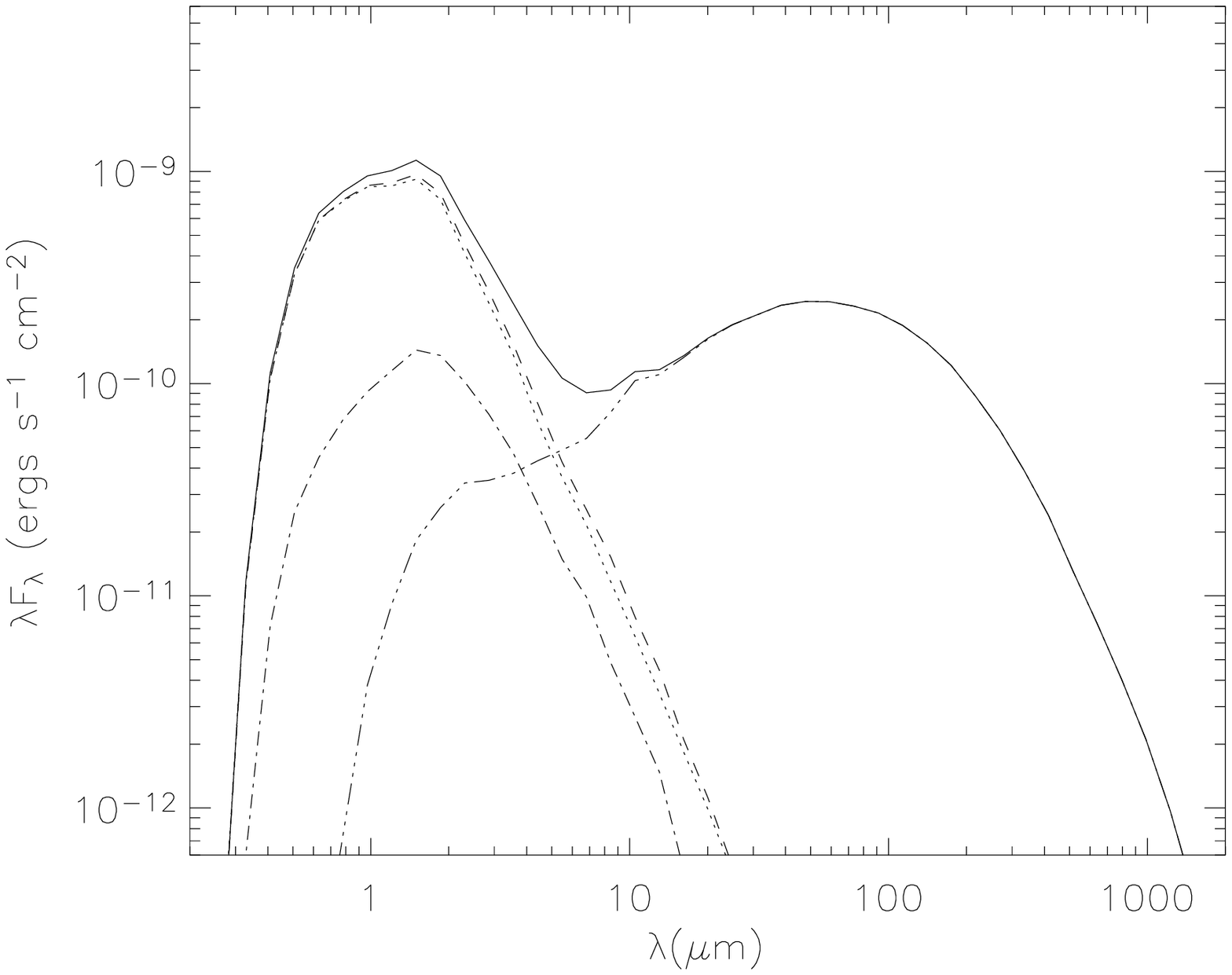,angle=0,width=2.5in}}
\caption{SEDs showing contributions from direct (dashed line), scattered 
(dot-dashed line) and disk reprocessed photons (triple dot-dash line) 
along with the input stellar spectrum (dotted line) and 
the total SED (solid line). Upper: brown dwarf model with 
$M_\star=0.01M_\odot$, log($M_d/M_\star$) = -1 and 
$L_\star=0.0013L_\odot$.  Lower: CTTS model with log($M_d/M_\star$)=-1}
\end{figure}

\newpage

\begin{figure}
\centerline{\psfig{figure=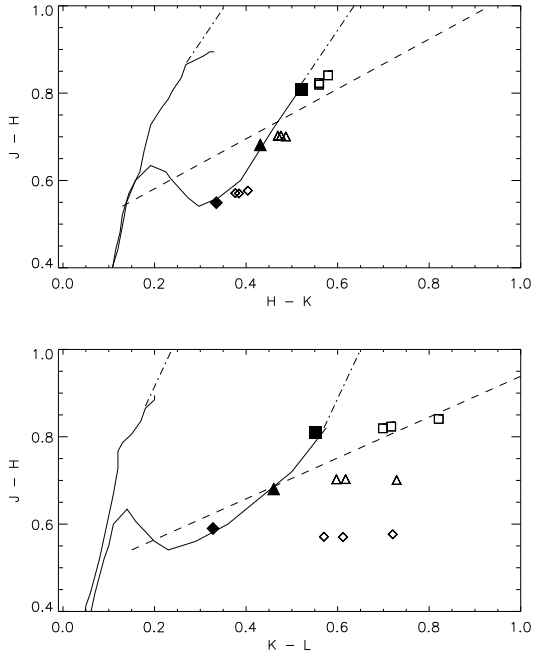,angle=0,width=3.in}}
\caption{JHKL color-color diagrams containing face-on colors 
for brown dwarf models with different $T_\star$, $M_d$ and $M_\star$.  
The reddest colors correspond to low stellar 
masses and therefore highly flared disks.  The filled symbols of 
corresponding shape represent the average stellar color of all models 
at fixed temperatures of 2200~K, 2600~K and 2800~K.  
The CTTS locus (dashed line) is taken from Meyer et al. (1997).  
The giant branch and M dwarf locus (solid lines) are taken 
from Bessell \& Brett (1988) and Kirkpatrick et al. (2000).  
The dot-dash lines represent reddening vectors. Models have 
been shifted so that average stellar colors lie on the ``brown dwarf locus''.}
\end{figure}

\newpage

\begin{figure}
\vspace{0.5cm}
\centerline{\psfig{figure=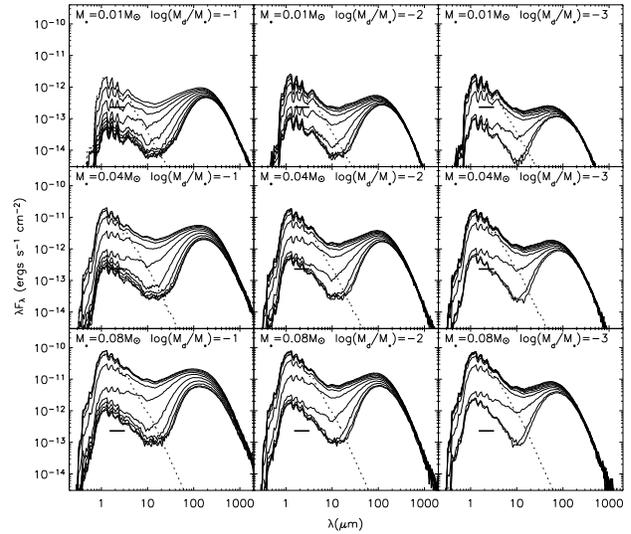,angle=0,width=3.5in}}
\caption{SEDs illustrating obscured fractions for our grid of brown 
dwarf models.  Obscured fractions range from $20\%$ up to $60\%$ whereas 
comparable CTTS models show 20\%.  The greatest obscuration in a brown 
dwarf model occurs for the lowest stellar mass of $M_\star=0.01M_\odot$, with 
log($M_d/M_\star$)=-1 and $T_\star=2200$~K/$L_\star=0.0013$~$L_\odot$.  
Least obscuration occurs for the highest stellar mass of 
$M_\star=0.08M_\odot$, with log($M_d/M_\star$)=-3 and 
$T_\star=2800$~K/$L_\star=0.044$~$L_\odot$.  
The thick horizontal solid line represents a detection limit of 16.5~mags 
at K.} 
\end{figure}

\newpage

\begin{figure}
\centerline{\psfig{figure=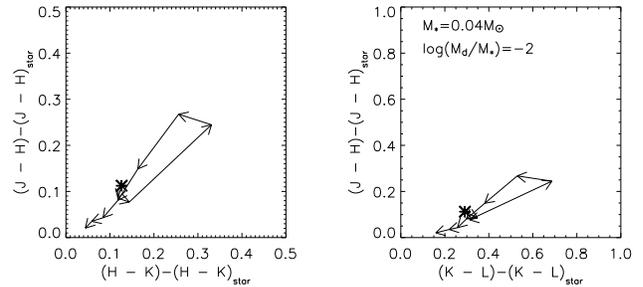,angle=0,width=3.5in}}
\caption{JHKL plot for the $M_\star=0.04M_\odot$ and log($M_d/M_\star$)=-2 
brown dwarf model. Arrows indicate the change in colors as inclination 
varies from nearly edge-on (indicated by bold asterisk) to face-on and 
colors are relative to central star's colors. This plot is indicative of 
the behaviour of all the brown dwarf models.}
\end{figure}

\newpage

\begin{figure}
\centerline{\psfig{figure=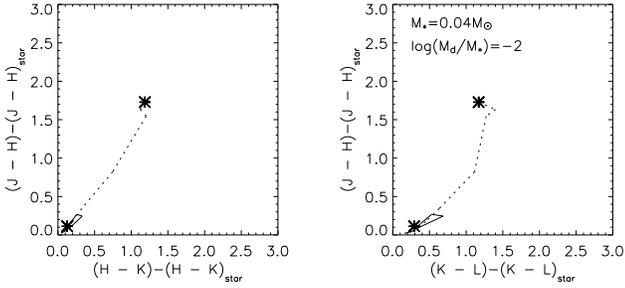,angle=0,width=3.5in}}
\caption{JHKL plots as in Fig.~7, but also with dotted lines 
showing the change 
at each inclination that would result if scattering effects were removed.}
\end{figure}

\newpage

\begin{figure}
\centerline{\psfig{figure=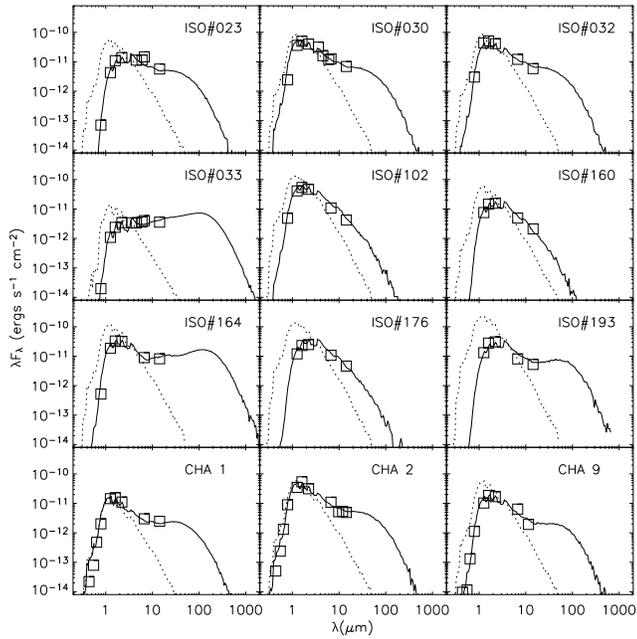,angle=0,width=3.5in}}
\caption{Models SED fits to observed data in $\rho$\ Ophiucus and Chameleon 
star clusters as indicated. All models are of a flared disk around a central 
star with an inner gap of radius 2$R_\star$.  
A distance of 150pc is assumed for both clusters.  
Further model parameters are given in Table.~2.  
The dotted line represents the input spectrum.  For ISO\#030 two models 
are presented.  In all cases a solid line indicates surface density 
$\Sigma\sim \varpi^{-1}$.  For ISO\#030 the dashed line indicates 
$\Sigma\sim \varpi^{-2}$.  The data is not corrected for 
the effects of reddening, instead the models are reddened 
according to Cardelli, Clayton and Mathis (1989) with $R_v=4$.}
\end{figure}

\newpage

\begin{figure}
\centerline{\psfig{figure=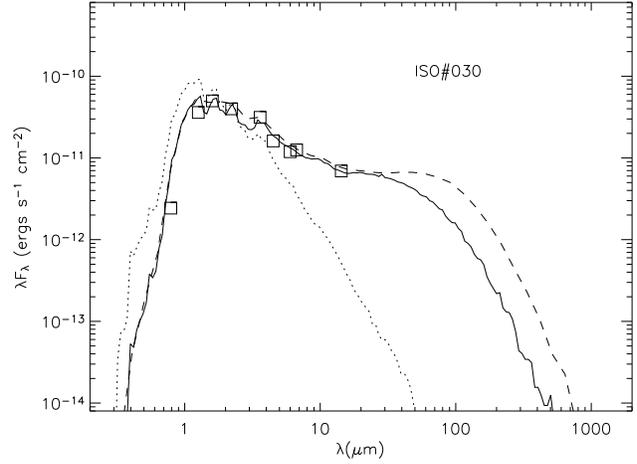,angle=0,width=3.5in}}
\caption{Models SED fits as in Fig9. Two models 
are presented for ISO\#030.  The solid line indicates surface density 
$\Sigma\sim \varpi^{-1}$.  The dashed line indicates 
$\Sigma\sim \varpi^{-2}$. Model parameters for $\Sigma\sim \varpi^{-1}$ 
case are given in Table.~2. The $\Sigma\sim \varpi^{-2}$ model has a 
disk eight time more massive.}
\end{figure}

\newpage

\begin{figure}
\centerline{\psfig{figure=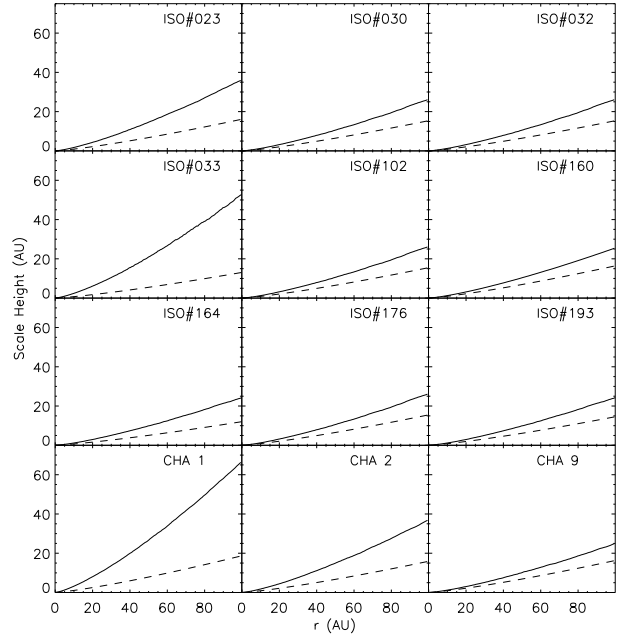,angle=0,width=3.5in}}
\caption{Scaleheights for our models (solid lines) 
presented in Fig.~9.  The scaleheight for our 
'typical CTTS' at matching disk to stellar mass ratio is 
given as a comparison (dashed line).  For ISO\#030 scaleheights 
are virtually coincident for the two models presented in Fig.~10 and 
therefore only the $\Sigma\sim \varpi^{-1}$ case is shown here.}
\end{figure}

\newpage

\begin{figure}
\centerline{\psfig{figure=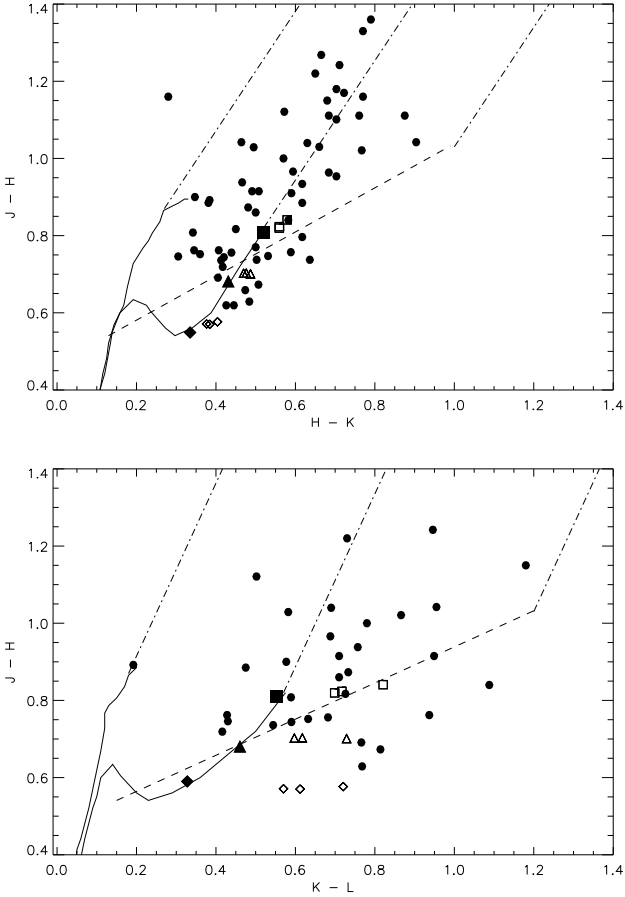,angle=90,width=3.5in}}
\caption{JHKL plots containing observed ($\bullet$) and face-on model 
(square,triangle,diamond) colors. 
Where possible data was obtained directly from the 2MASS second 
incremental data release (Carpenter, 2001).  Remaining data was 
retrieved from the papers of Comer\'{o}n et al (1998/2000), 
Mart\'{i}n et al (2000), Brice\~{n}o et al (1998), Luhman et al (1998a), 
Luhman (1999), Najita et al (2000) and  Liu et al (2003).  
As previously the solid shapes represent the average stellar color 
obtained for models of $T_\star=$2200, 2600 and 2800~K.  
The CTTS locus (dashed line) is taken from Meyer et al. (1997).  
The giant branch and M dwarf locus (solid lines) are taken from 
Bessell \& Brett (1988) and Kirkpatrick et al. (2000).  
The dot-dash lines represent reddening vectors.  
Models have been shifted so that stellar colors lie on the 
``brown dwarf locus''.}
\end{figure}

\newpage

\begin{figure}
\centerline{\psfig{figure=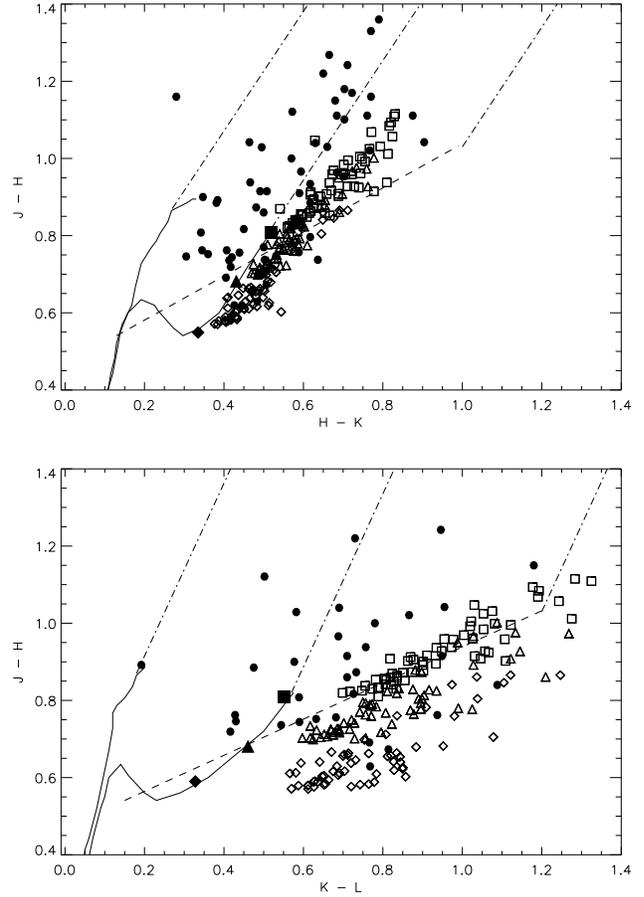,angle=90,width=3.5in}}
\caption{JHKL plots as in Fig.~12, but with model colors for all inclinations.}
\end{figure}

\newpage


\begin{table}
\caption{Model Parameters}
\begin{tabular}{@{}cccc}
\hline
\hline
\\
$M_\star$&$T_\star$&$R_\star$&$L_\star$\\
($M_\odot$)&(K)&($R_\odot$)&($L_\odot$)\\
\\
\hline
\\
0.01&2200&0.25&0.0013\\
0.04&2600&0.50&0.0038\\
0.08&2800&0.90&0.044\\
\\
\hline
\end{tabular}
\end{table}


\begin{table}
\caption{Model Fit Parameters}
\begin{tabular}{@{}ccccccc}
\hline
\hline
\\
Object&$T_\star$&$R_\star$&$M_\star$&$M_d$&$A_v$&Inclination\\
&(K)&($R_\odot$)&($M_\odot$)&($M_\odot$)&(mags)&(deg)\\
\\
\hline
\\
ISO\#023&2600&0.95&0.04&$10^{-5}$&8&0\\
ISO\#030&2600&1.2&0.08&$10^{-5}$&2&0\\
ISO\#032&2600&1.2&0.08&$10^{-5}$&3&0\\
ISO\#033&2200&0.63&0.01&$10^{-3}$&7&0\\
ISO\#102&3000&1.17&0.08&$10^{-7}$&3.5&0\\
ISO\#160&2600&0.95&0.08&$10^{-7}$&6&0\\
ISO\#164&2600&1.36&0.08&$10^{-3}$&4&63\\
ISO\#176&3000&1.17&0.08&$10^{-7}$&7&60\\
ISO$\#$193&3000&1.5&0.08&$10^{-5}$&7&78\\
CHA H$\alpha$1&2600&0.5&0.01&$10^{-5}$&0.3&0\\
CHA H$\alpha$2&2600&1.05&0.04&$10^{-5}$&1.1&37\\
CHA H$\alpha$9&2600&0.95&0.08&$10^{-5}$&3.2&72
\\
\hline
\end{tabular}
\end{table}

\label{lastpage}

\end{document}